\documentclass[conference]{IEEEtran}
\IEEEoverridecommandlockouts
\usepackage{cite}
\usepackage{amsmath,amssymb,amsfonts}
\usepackage{algorithmic}
\usepackage{graphicx}
\usepackage{textcomp}
\usepackage[numbers]{natbib}  

\usepackage{subfig}
\usepackage[table,xcdraw]{xcolor} 
\usepackage{floatrow}
\usepackage{multirow}

\usepackage{xcolor}
\usepackage{url}

\usepackage{soul} 

\usepackage{refcount} %

\usepackage{lscape}

\usepackage{todonotes}

\def\BibTeX{{\rm B\kern-.05em{\sc i\kern-.025em b}\kern-.08em
    T\kern-.1667em\lower.7ex\hbox{E}\kern-.125emX}}

\begin{document}

\title{A Taxonomy for Blockchain-based Decentralized Physical Infrastructure Networks (DePIN)
\thanks{M.~Ballandies is supported by the European Union's Horizon 2020 research and innovation programme under grant agreement No 833168.}
}

\author{\IEEEauthorblockN{Mark C. Ballandies}
	\IEEEauthorblockA{
        Computational Social Science \\
		ETH Zurich\\
		{\small bcmark@protonmail.com} }
  \\
  \IEEEauthorblockN{Joshua C. Yang}
	\IEEEauthorblockA{Computational Social Science\\
		ETH Zurich \\
  {\small joshua.yang@gess.ethz.ch} }
  \and
  \IEEEauthorblockN{Hongyang Wang}
	\IEEEauthorblockA{Civil Engineering \\
        ETH Zurich \\
		{\small wang@ibi.baug.ethz.ch} }
  \\
  \IEEEauthorblockN{Christophe Gösken}
	\IEEEauthorblockA{Center for Law \& Economics\\
		ETH Zurich \\
       {\small christophe.goesken@gess.ethz.ch}
  } 
  \and
  \IEEEauthorblockN{Andrew Chung Chee Law}
	\IEEEauthorblockA{Research Scientist\\
        IoTeX \\
		{\small andrew@iotex.io} }
  \\
  \IEEEauthorblockN{Michael Andrew}
	\IEEEauthorblockA{Project Initiator\\
		Who Loves Burrito? \\
          {\small w.l.burrito@wholovesburrito.com } }

}

\maketitle

\begin{abstract}
As digitalization and technological advancements continue to shape the infrastructure landscape, the emergence of blockchain-based decentralized physical infrastructure networks (DePINs) has gained prominence. However, a systematic categorization of DePIN components and their interrelationships is still missing. To address this gap, we conduct a literature review and analysis of existing frameworks and derived a taxonomy of DePIN systems from a conceptual architecture. Our taxonomy encompasses three key dimensions: distributed ledger technology, cryptoeconomic design and physicial infrastructure network. Within each dimension, we identify and define relevant components and attributes, establishing a clear hierarchical structure. Moreover, we illustrate the relationships and dependencies among the identified components, highlighting the interplay between governance models, hardware architectures, networking protocols, token mechanisms, and distributed ledger technologies. This taxonomy provides a foundation for understanding and classifying diverse DePIN networks, serving as a basis for future research and facilitating knowledge exchange, fostering collaboration and standardization within the emerging field of decentralized physical infrastructure networks.
\end{abstract}

%

\section{Introduction}

Traditional industries of physical infrastructures, such as telecommunications, 
satellite radio, 
and cartography, 
have operated under centralized control with single entities or associations of a few centralized entities of predominant authority. With the advancement of the internet and communication technologies, many infrastructure ecosystems instigated a decentralized trend against traditional centralization. However, in recent times, these formerly open and inclusive digitized platforms and services have increasingly shifted the focus towards prioritizing their private interests. Such is the case of the cartography industry, dominated by a single service provider, Google Maps which has gradually re-centralized the industry~\cite{plantin2018digital}.

With blockchain technology, however, it has become possible to decentralize the physical networks of infrastructure. Through the implementation of automatized operation, immutable common data storage and redesigned value-sensitive governance, the potential arises for distributing control, ownership, and decision-making among multiple participants or nodes in a network \cite{pieroni2020blockchain,roman2013features,wang2022if, helbing2023democracy}. Decentralized physical infrastructure networks (DePINs) are an emerging trend within blockchain \cite{kassab2023sectormap} that describes this growing ecosystem of web3 projects that utilize token incentives, smart contracts, decentralization, and participatory governance mechanisms of blockchains to deploy networks of real devices such as sensors, storage, or wireless networks globally. 

Decentralization of physical networks has been recognized to result in several benefits including increased resilience~\cite{roman2013features}, due to distributing the system away from a single point of failure and system redundancy; trust ~\cite{de2020blockchain}, due to data transparency and tamper-proof guarantees; and accessibility~\cite{nakamoto_bitcoin_2008} due to permissionless requirements of joining the network. Additionally, the operation of infrastructure can be improved due to the re-structure of incentive mechanisms (e.g. via blockchain-based tokens), flexible decentralized marketplaces, and inclusive participation~\cite{dapp2021finance}.

At the time of writing, more than 50 blockchain systems can be counted in this ecosystem with varying designs, layouts and application domains \cite{burrito}. Nevertheless, a classification of these systems based on a rigorously from theory derived taxonomy is still missing. Such a taxonomy would facilitate the differentiation and comparison among systems and enable researchers to identify unexplored system layouts and practitioners to learn from others, thus facilitating innovation. In particular, the lack of a classfication of DePIN systems on a rigorous taxonomy can result in a fragmentation of the community and thus duplication of efforts \cite{ballandies2022decrypting}. 



This work contributes a first stepping stone to conduct such a comprehensive classification with the following contributions:

\begin{enumerate}
    \item A general conceptual architecture illustrating blockchain-based systems
    \item A taxonomy for DePIN derived from this conceptual architecture
\end{enumerate}


\section{Literature Review}
\label{sec:lit_review}
 Decentralized physical infrastructure networks (DePINs) are cryptoeconomic systems. These are systems that consist of i) individual autonomous actors, ii) economic policies embedded in software and iii) emergent properties arising from the interactions of those actors according to the rules defined by that software" \cite{ballandies2022fundamentals}. In the case of DePIN, autonomous actors are incentivized to place or use physical devices on a global scale via mechanisms enabled by distributed ledger technology~(DLT) (as utilized in blockchain) from which a physical infrastructure may arise as an emergent property.

Since such DLT-based cryptoeconomic systems are complex systems \cite{ballandies2022fundamentals} that are difficult to control and govern hierarchically from the top down due to their emergent nature, decentralized mechanisms have been proposed and are used in DePIN: i) For control, DLT-based tokens are used as an incentive mechanism. In particular, such tokens can align human behavior with goals set by a community  \cite{dapp2021finance} (e.g., the establishment of a DePIN). ii) For governance, shared ownership of the network can be achieved via token-based decision-making mechanisms on system parameters, compositions and general vision (e.g. via improvement proposals).  Usually, all token holders can participate in these processes~\cite{wang2022if,machart_governance_2020}. 

In particular, hierarchical and centralized governance mechanisms are often challenged in complex systems \cite{helbing2023democracy} and decentralized and bottom-up mechanisms are required to govern and control them. For instance, these systems can be efficiently governed via decentralized mechanisms such as collective intelligence, digital democracy and self-organization \cite{helbing2023democracy} which hence found not only a natural expression globally in urban participatory budgeting projects\footnote{Switzerland's city of Aarau Participatory Budgeting: \url{https://www.stadtidee.aarau.ch/abstimmungsphase.html/1937}} \citep{buterin2022proof}, or self-organizing business teams \cite{laloux2014reinventing}, but also in blockchain-based web3 systems in the form of decentralized autonomous organizations (DAOs). These DAOs utilize the collective intelligence of their community (e.g. via open discussion platforms) to identify ideas and then supports them to deliberate (e.g. via Improvement proposals), decide (e.g. via token-based voting) and implement (e.g. steered/ controlled via token-based incentives) them effectively \cite{merkle_daos_nodate}. 
Often, these mechanisms are implemented using smart contracts. In general, this DLT-based code can improve the functioning of society by automating and increasing the transparency of the implemented mechanisms  \cite{heckler2022crypto}. 

DePIN represents a subset of the broader blockchain IoT domain, characterized by its utilization of physical hardware or resources to deliver tangible or digital services to consumers. This subset opens up new possibilities and applications within the blockchain IoT framework, driving the emergence of DePIN projects that cater to specific use cases in the realm of physical service provisioning.

Because DePIN operates at this convergence of blockchain, Internet-of-Things (IoT), and physical service delivery, certain aspects of it can be described with existing taxonomies in the field of blockchain IoT \cite{abdelmaboud2022blockchain}. Notably, i) the underlying hardware architecture in DePIN projects shares similarities with the utilization of  hardware in acquiring data or providing fungible or non-fungible goods and services, akin to the perception or sensor layer observed in traditional Blockchain IoT systems~\cite{pieroni2020blockchain}. Moreover ii), drawing parallels with off-chain compute in the context of blockchain IoT systems \cite{leng2022secure}, the middleware plays a pivotal role in processing, storing, transporting data, and relaying services acquired by the hardware layer in DePIN projects. 

The establishment of a comprehensive taxonomy is of paramount importance in providing clear guidelines for stakeholders aiming to foster the growth of emerging sectors. While the Internet of Things (IoT) taxonomy has been thoroughly explored in the literature \cite{yaqoob_internet_2017}, and the advent of blockchain and blockchain IoT has gained substantial popularity, resulting in the emergence of general DLT taxonomies (see \citep{ballandies2022decrypting} for an overview) and fewer blockchain IoT works \cite{gurpinar2022blockchain, siegfried_blockchain_2022}, due to its recent history and distinct characteristics focus on instantiating infrastructure networks, there remains a conspicuous absence of well-defined taxonomy for DePIN that combines previously introduced taxonomies from DLT and IoT and extends them to be applicable for infrastructure networks. 

In order to close this gap, this work introduces a rigorously derived taxonomy for DePIN that considers the cryptoeconomic design, distributed ledger technology and physical infrastructure components of DePINs.

\begin{table*}[]
\begin{tabular}{|l|l|l|l|}
\hline
\textbf{Location}                        & \textbf{Component}                        & \textbf{Definition}                                                                                                                                                                                                                                    & \textbf{Chain}                       \\ \hline
                                         & \cellcolor[HTML]{EFEFEF}Governance action & \begin{tabular}[c]{@{}l@{}}A governance action is one or more real-life activities connected to\\ governance that can be digitally represented in a DLT system\end{tabular}                                                                            &                                      \\ \cline{2-3}
                                         & Core action                               & \begin{tabular}[c]{@{}l@{}}A core action is one or more real-life activities connected to the\\ economy of the system that can be digitally represented in a DLT \\ system.\end{tabular}                                                               &                                      \\ \cline{2-3}
\multirow{-3}{*}{\textit{Real World}}    & \cellcolor[HTML]{EFEFEF}Hardware          & \begin{tabular}[c]{@{}l@{}}Hardware are physical devices required to participate in the\\ (core economy of the) DLT system.\end{tabular}                                                                                                               &                                      \\ \cline{1-3}
                                         & \cellcolor[HTML]{EFEFEF}Middleware        & \begin{tabular}[c]{@{}l@{}}Middleware are services such as storage, communication, or messaging\\ between components within a DLT system that are not covered by DLT.\end{tabular}                                                                     &                                      \\ \cline{2-3}
                                         & Consensus                                 & \begin{tabular}[c]{@{}l@{}}Consensus is the mechanism through which entries are written to the\\ distributed ledger, while adhering to a set of rules that all participants\\ enforce when an entry containing transactions is validated.\end{tabular} & \multirow{-5}{*}{\textit{off-chain}} \\ \cline{2-4} 
                                         & DL                                        & \begin{tabular}[c]{@{}l@{}}A distributed ledger is defined as a distributed data structure,\\ containing entries that serve as digital records of actions.\end{tabular}                                                                                &                                      \\ \cline{2-3}
                                         & \cellcolor[HTML]{EFEFEF}Smart contracts   & \begin{tabular}[c]{@{}l@{}}Smart contracts are code or mechanisms that are deployed to the  \\ distributed ledger and executed by a DLT system.\end{tabular}                                                                                                                                  &                                      \\ \cline{2-3}
\multirow{-5}{*}{\textit{Digital World}} &Token             & \begin{tabular}[c]{@{}l@{}}Token is a unit of value issued within a DLT system and which can\\ be used as a medium of exchange or unit of account.\end{tabular}                                                                                        & \multirow{-3}{*}{\textit{on-chain}}  \\ \hline
\end{tabular}
 \caption{Components of the conceptual architecture for DePIN modeled on \citet{ballandies2022decrypting}. Shown in gray are the components that were added to account for DePIN.}\label{fig:token_dilution}
\end{table*}



\section{Methodology}

This work follows the method of \citet{nickerson2013method} for deriving a taxonomy for DePIN and performs one iteration of the \textit{Conceptual-to-empirical} step: 1.  An established conceptual architecture and taxonomy of Blockchain systems \citep{ballandies2022decrypting} is extended by theoretical reasoning considering the meta-characteristic "Functioning of decentralized physical infrastructure networks". 2. DePIN systems are classified based on that taxonomy and 3. The taxonomy is revised.

\section{Conceptual Architecture}
\label{sec:conceptual_architecture}
Considering the meta-characteristic and the exploration in Section \ref{sec:lit_review}, the conceptual architecture found in literature \citep{ballandies2022decrypting} is extended such that it illustrates the inner mechanisms of DePIN and the interrelationships of their components. 

Because DePIN are about physical infrastructures, both a hardware and a middleware component are added to the conceptual architecture. 

Further, as governance is an important component of DLT systems 
\citep{zwitter2020decentralized}, the governance component has been integrated into the conceptual architecture. In particular, the original Action component has been renamed to \textit{core economy action} to better separate the economic and governance aspects of a DePIN system. 

Finally, to make the implemented mechanisms of DePIN such as tokens and DAOs more transparent and decentralized, on-chain features of DLT such as smart contracts are often used. Therefore, the Smart Contracts component was added to illustrate which parts of a DePIN are secured and supported by DLT.

\section{Taxonomy}

\begin{figure*}[tb!]
\begin{center}
\includegraphics[width=0.9\textwidth]{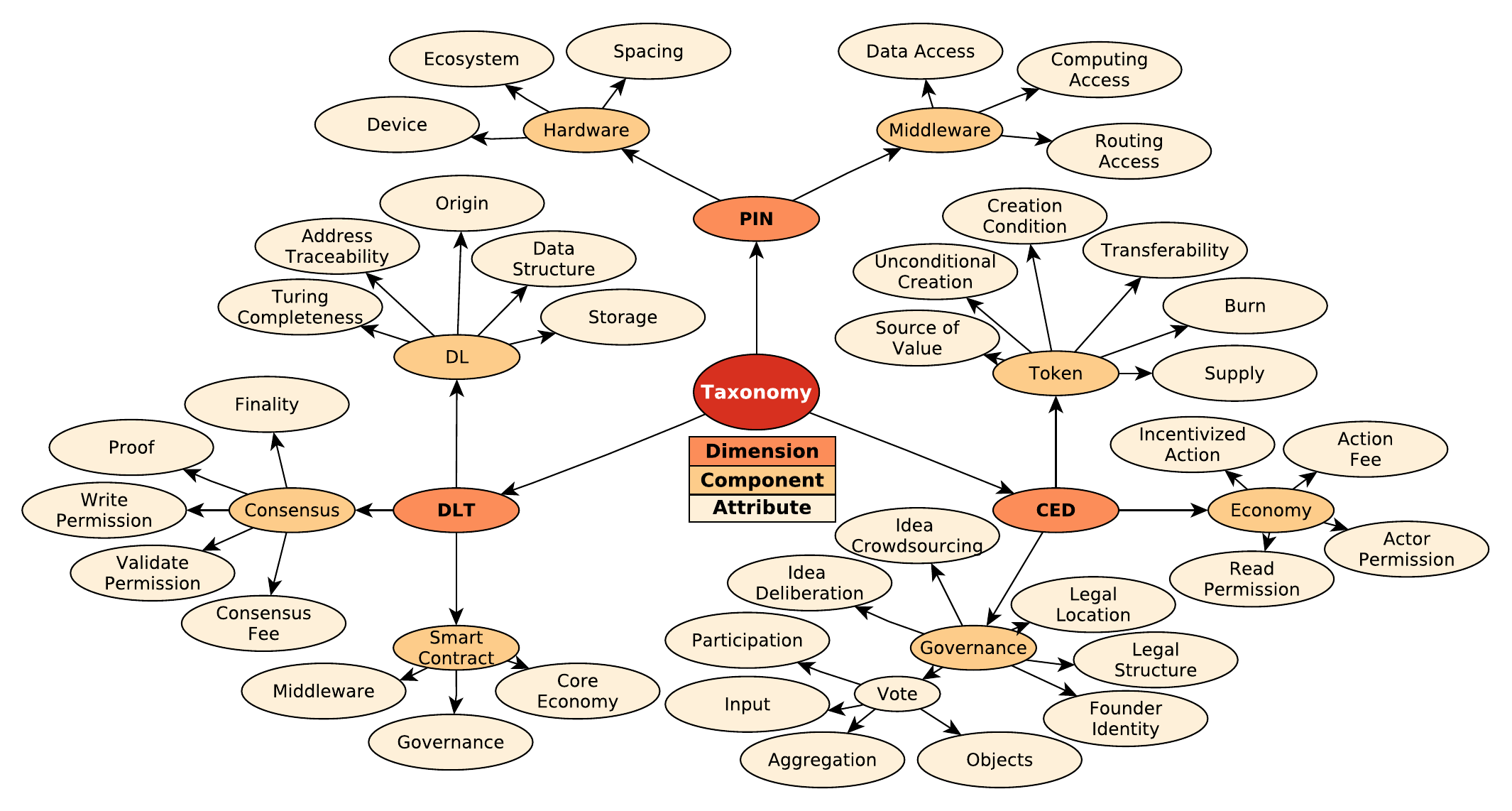} 
\end{center}
\caption{The taxonomy with its three dimensions (distributed ledger technology (DLT), pysicial infrastructure network (PIN), and cryptoeconomic design (CED)), 8 components and 41 attributes.} \label{fig:taxonomy}
\end{figure*}

Based on the conceptual architecture (Section \ref{sec:conceptual_architecture}), further components and attributes are added to the Taxonomy introduced in \citep{ballandies2022decrypting}. The taxonomy is illustrated in Figure \ref{fig:taxonomy} and the main differences are explained below.

\subsection{Hardware}

The hardware component, which represents the hardware used in the DLT system in more detail, has been added to the new Physical Infrastructure Network (PIN) dimension of the taxonomy (Figure \ref{fig:taxonomy}).
\textbf{Device} illustrates what hardware devices system participants are required to obtain in order to participate in the core economy of the network. The characteristics can be in the context of DePIN for instance \textit{camera} (e.g. HiveMapper or NATIX) or \textit{LoRaWAN hotspot} (e.g. Helium). 
\textbf{Ecosystem} illustrates the degree of openness of the hardware ecosystem. The characteristic is \textit{open} if any manufacturers devices can be utilized to participate in the core economy (e.g. onocoy or WiHi) or \textit{licensed} if a manufacturer is required to get a license before being able to produce devices for the system (e.g. Helium) or \textit{closed} if manufacturing is restricted to specific manufacturers  (e.g. GEODNET).
\textbf{Spacing} illustrates if devices are required to have a minimum distance from each other. The characteristic is \text{Yes} if there are measures in place to facilitate a particular spacing between devices and \textit{No} if devices can be placed anywhere. For instance, in the case of DePINs in the field of global navigation satellite systems such as onocoy or GEODNET, overpopulation of hardware devices in a given area is not desirable as it does not bring any further benefit to the network, while in the case of the Render network, which builds a decentralized GPU infrastructure, such spacing is not required. 


\subsection{Middleware}
The middleware component, which illustrates the middleware used in the DLT system that is not executed on-chain (e.g., via smart contracts), was added to the new Physical Infrastructure Network (PIN) dimension (Figure \ref{fig:taxonomy}).
\textbf{Data access} illustrates the degree of decentralization in the storage and access of system data required for the core economy and governance. The characteristics are \textit{open} if the public can participate in storing and accessing the data important for the systems functioning or \textit{restricted} if only selected entities can provide this service. For instance, solutions such as Filecoin \cite{psaras2020interplanetary} can facilitate the storing of system information in a decentralized way in a DePIN system. 
\textbf{Routing access} illustrates the degree of decentralization in the routing of data between system participants/ devices of the core economy and governance. The characteristics are \textit{open} if the public can participate in the routing of information important for the systems' functioning or \textit{restricted} if only selected entities can route information. For instance, solutions such as W3bstream~\cite{fan2023connecting} or Streamr \cite{savolainen2020streamr} can be utilized by a DePIN to facilitate the routing of information in a decentralized way. 
\textbf{Computing access} illustrates the degree of decentralization in the computation of system mechanisms/ quantities of the core economy and governance.

\subsection{Core economy action}
The core economy component is extended with one attribute.
\textbf{Incentivized action}
Illustrates if and which of the actions taken in the core economy is incentivized with the awarding of token units. The characteristics can be for instance in the DePIN context \textit{hardware} placement, but in a more general context also \textit{staking}, or \textit{system parameter calibration}.

\subsection{Governance}
The governance component is added to the cryptoeconomic design (CED) of the taxonomy.

\textbf{Idea crowd-sourcing} illustrates who can express ideas in system channels that can be used in the deliberation process and thus represents the collective intelligence of a DePIN. 
The characteristics can be \textit{public} or \textit{restricted}. 
\textbf{Idea deliberation} illustrates who can create, propose, curate, and merge voteable items.
\textbf{Vote objects} identifies on what system participants can vote. The characteristics are \textit{system parameters}, \textit{improvement proposals}, \textit{governance bodies}, a combination of the previous, or \textit{none}. For instance, in DePIN systems, often token holders can vote on improvement proposals (e.g. Helium, DIMO or Render). In some DeFi systems (e.g. MakerDAO) token holders can also directly vote on system parameter configurations. 
\textbf{Vote participation} illustrates who can participate in the voting. It is \textit{public} if anyone/ any Token holder can vote or \textit{restricted} if a voter is required to fulfill further requirements such as know-your-customer policies besides owning tokens and \textit{closed} if voting is not accessible for token holders.  
\textbf{Voting input} illustrates how the votes can be cast by an address or unique identity. The feature can be \textit{Single choice} (e.g. utilized in Helium), if the voters can select one option. \textit{Multiple-selection} if voters can select several options (e.g. utilized in DIMO), or 
\textit{Ranked-choice} (e.g. utilized in DIMO), if the voters rank the different choices according to their preference; \textit{Quadratic} \citep{lalley2018quadratic} (e.g. modified version utilized in onocoy), 
when voters use credits to vote for options, but the cost of casting multiple votes for the same option increases quadratically. 
\textbf{Voting aggregation} illustrates the methods of aggregation. This is the calculation of votes after the action of voting. These methods may involve amongst other \textit{Simple Majority} ($> 50\%$) or \textit{Super Majority} ($> 66\%$, e.g. as in Helium).
\textbf{Founder/ core-team identity} illustrates the degree to which the identity of the system initiators and implementors are known. 
\textbf{Legal structure} illustrates the legal interpretation of the governance components, according to international, regional and/or national laws and regulations. For instance, often DePIN system utilize a Foundation as the IP-owning entity (e.g. Helium, DIMO, Render). Nevertheless, also other setups exists, such as in the case of onocoy where an association is chosen to be the IP-owning entity. 
\textbf{Location} illustrates where the DePIN operates and which laws and regulations apply to it.

\subsection{Smart contracts}
The smart contract component has been added to the DLT dimension of the taxonomy. 
[\textbf{Governance, Core economy, Middleware}] illustrates the degree to which the [\textit{governance, core economy, middleware}] of the system is implemented on-chain. The charactersitic is \textit{all} if the whole, \textit{partial} if a part, and \textit{none} if no [\textit{governance, core economy, middleware}] is implemented on-chain.  




\section{Findings}


Through our efforts to classify DePIN projects using the proposed taxonomy, four observations are obtained:

First, the analysis shed light on the significant variations in the level of decentralization within the governance design of DePIN projects. It became evident that some projects exhibited a centralized approach, where decision-making authority predominantly resides with a hierarchically organized core team, while in others a self-organizing community participates actively in decision-making processes through diverse voting mechanisms. Mostly this open governance participations in DePIN systems is restricted to improvements proposals which are not legally binding and thus neither impact system parameters nor the governance body/ workings of IP-owing legal entities directly. 

Second, though there are solutions with for instance W3bstream \cite{fan2023connecting}, Streamr \cite{savolainen2020streamr}, Filecoin \cite{psaras2020interplanetary} and others to decentralize off-chain middleware, the middleware in DePIN systems is currently often centralized (maintained by centralized entities and deployed to centralized cloud architectures). 

Third, DePIN projects are inherently complex, encompassing multifaceted elements. Thus, acquiring comprehensive information is a challenging endeavour. Different DePIN projects showcased different levels of openness and accessibility of information. Information pertaining to governance structures, software and hardware technical details, as well as investor information and fundraising mechanisms, were often difficult to obtain. This lack of transparency and availability of information poses a notable hurdle in comprehensively understanding and evaluating DePIN projects. For instance, initially a \textit{investor identity} attribute had been added to the Governance component, but was removed because of the inability to clearly assign a value for it for classified systems.

Fourth, it can be a challenge to measure attributes of a DePIN or DLT system quantitatively. For instance, the taxonomy had the \textit{information access} attribute to illustrate how open the information regarding the system (for instance, including the internal strategies of the core team) are, resp. who has access to them. For instance, more transparency/ openness would indicate an improved collective intelligence~\cite{helbing2023democracy} of the community and thus would be important to be classified.

Besides these observations, it was noticed that the extension of the conceptual architecture/ taxonomy introduced in \citet{ballandies_decrypting_2022} with the new components and attributes of this work did not restrict the taxonomy’s explanatory capability to illustrate non-DePIN related DLT systems. For instance, the \textit{hardware device} in the Bitcoin system are ASICS and the \textit{incentivized action} is proof-of-work participation. 

Finally, the term 'DePIN' has been used widely to refer to decentralized token-incentivized blockchain networks or systems, however, more recently, there has been a push to clearly delineate between physical networks and strictly digital resources such as 'computational resources'. We found that the taxonomy can clearly illustrate the difference between DePIN and non-DePIN systems: Whenever the incentivized action in a system is the placement or contribution of physical infrastructure elements to the system, such as a camera or storage drives, it can be defined as a DePIN system.


\section{Conclusion and outlook}

This taxonomy paper has presented a comprehensive and novel framework for classifying DePIN based on cryptoeconomic design, distributed ledger technologies and physical infrastructure network dimensions that span 8 components ranging from tokens over governace to consensus and hardware. It intends to provide a valuable foundation for understanding and categorizing DePIN. 

Further research could focus on addressing the challenges associated with information availability and complexity within the DePIN domain. This could include initiatives to develop standardized reporting frameworks, guidelines for information disclosure, and collaborative efforts to create comprehensive databases of DePIN projects. By addressing these issues, the understanding and evaluation of DePIN projects can be significantly enhanced, contributing to the broader development and adoption of decentralized physical infrastructure networks. Beyond infrastructure networks, the broader context of the built environment, including smart cities and intelligent buildings, holds significant potential for further DePIN research. 
For instance, integrating DePIN with artificial intelligence (AI), creates opportunities for autonomous, and sovereign infrastructure and buildings \cite{hunhevicz2021no1s1}.

Additionally, this taxonomy framework has the versatility to be applied to other vertical studies within the broader domain of decentralized infrastructure. It offers opportunities for cross-disciplinary collaboration and facilitates knowledge sharing across different sectors, such as value-sensitive governance design, organizational structure studies, complex systems, legal and regulatory studies, and other vertical theoretical explorations. In particular, a rigorous crowd-sourced classification, analysis and validation can further enhance the taxonomy and derive general design patterns~\cite{ballandies2022decrypting}.

\bibliographystyle{unsrtnat}
\bibliography{taxonomy,Depin_una}

\begin{thebibliography}{30}
\providecommand{\natexlab}[1]{#1}
\providecommand{\url}[1]{\texttt{#1}}
\expandafter\ifx\csname urlstyle\endcsname\relax
  \providecommand{\doi}[1]{doi: #1}\else
  \providecommand{\doi}{doi: \begingroup \urlstyle{rm}\Url}\fi

\bibitem[Plantin(2018)]{plantin2018digital}
Jean-Christophe Plantin.
\newblock Digital traces in context| google maps as cartographic
  infrastructure: From participatory mapmaking to database maintenance.
\newblock \emph{International journal of communication}, 12:\penalty0 18, 2018.

\bibitem[Pieroni et~al.(2020)Pieroni, Scarpato, and
  Felli]{pieroni2020blockchain}
Alessandra Pieroni, Noemi Scarpato, and Lorenzo Felli.
\newblock Blockchain and iot convergence—a systematic survey on technologies,
  protocols and security.
\newblock \emph{Applied Sciences}, 10\penalty0 (19):\penalty0 6749, 2020.

\bibitem[Roman et~al.(2013)Roman, Zhou, and Lopez]{roman2013features}
Rodrigo Roman, Jianying Zhou, and Javier Lopez.
\newblock On the features and challenges of security and privacy in distributed
  internet of things.
\newblock \emph{Computer networks}, 57\penalty0 (10):\penalty0 2266--2279,
  2013.

\bibitem[Wang et~al.(2022)Wang, Hunhevicz, and Hall]{wang2022if}
Hongyang Wang, Jens Hunhevicz, and Daniel Hall.
\newblock What if properties are owned by no one or everyone? foundation of
  blockchain enabled engineered ownership.
\newblock In \emph{EC3 Conference 2022}, volume~3, pages 0--0. University of
  Turin, 2022.

\bibitem[Helbing et~al.(2023)Helbing, Mahajan, Fricker, Musso, Hausladen,
  Carissimo, Carpentras, Stockinger, Sanchez-Vaquerizo, Yang,
  et~al.]{helbing2023democracy}
Dirk Helbing, Sachit Mahajan, Regula~H{\"a}nggli Fricker, Andrea Musso,
  Carina~I Hausladen, Cesare Carissimo, Dino Carpentras, Elisabeth Stockinger,
  Javier~Argota Sanchez-Vaquerizo, Joshua~C Yang, et~al.
\newblock Democracy by design: Perspectives for digitally assisted,
  participatory upgrades of society.
\newblock \emph{Journal of Computational Science}, 71:\penalty0 102061, 2023.

\bibitem[Kassab(2023)]{kassab2023sectormap}
Sami Kassab.
\newblock The depin sector map.
\newblock \url{https://messari.io/report/the-depin-sector-map}, 2023.
\newblock (Accessed on 03/07/2023).

\bibitem[De~Filippi et~al.(2020)De~Filippi, Mannan, and
  Reijers]{de2020blockchain}
Primavera De~Filippi, Morshed Mannan, and Wessel Reijers.
\newblock Blockchain as a confidence machine: The problem of trust \&
  challenges of governance.
\newblock \emph{Technology in Society}, 62:\penalty0 101284, 2020.

\bibitem[Nakamoto()]{nakamoto_bitcoin_2008}
Satoshi Nakamoto.
\newblock Bitcoin: A peer-to-peer electronic cash system.
\newblock page~9.

\bibitem[Dapp et~al.(2021)Dapp, Helbing, and Klauser]{dapp2021finance}
Marcus~M Dapp, Dirk Helbing, and Stefan Klauser.
\newblock \emph{Finance 4.0-Towards a Socio-Ecological Finance System: A
  Participatory Framework to Promote Sustainability}.
\newblock Springer Nature, 2021.

\bibitem[Andrew(2023)]{burrito}
Michael Andrew.
\newblock Depin project list, 2023.
\newblock URL \url{https://wholovesburrito.com/project-list/}.

\bibitem[Ballandies et~al.(2022)Ballandies, Dapp, and
  Pournaras]{ballandies2022decrypting}
Mark~C Ballandies, Marcus~M Dapp, and Evangelos Pournaras.
\newblock Decrypting distributed ledger design—taxonomy, classification and
  blockchain community evaluation.
\newblock \emph{Cluster computing}, 25\penalty0 (3):\penalty0 1817--1838, 2022.

\bibitem[Ballandies(2022)]{ballandies2022fundamentals}
Mark~Christopher Ballandies.
\newblock \emph{Fundamentals of Cryptoeconomics: On the design, construction,
  and impact of blockchain-based systems and incentives}.
\newblock PhD thesis, ETH Zurich, 2022.

\bibitem[Machart and Samadi()]{machart_governance_2020}
Felix Machart and Jascha Samadi.
\newblock Governance by and of blockchains.
\newblock page 115.

\bibitem[Buterin and Schneider(2022)]{buterin2022proof}
Vitalik Buterin and Nathan Schneider.
\newblock \emph{Proof of stake: the making of Ethereum, and the philosophy of
  blockchains}.
\newblock Seven Stories Press, 2022.

\bibitem[Laloux and Wilber(2014)]{laloux2014reinventing}
Frederic Laloux and Ken Wilber.
\newblock \emph{Reinventing organizations: A guide to creating organizations
  inspired by the next stage of human consciousness}, volume 360.
\newblock Nelson Parker Brussels, 2014.

\bibitem[Merkle()]{merkle_daos_nodate}
Ralph~C Merkle.
\newblock {DAOs}, democracy and governance.
\newblock page~28.

\bibitem[Heckler and Kim(2022)]{heckler2022crypto}
Nuri Heckler and Yeonkyung Kim.
\newblock Crypto-governance: The ethical implications of blockchain in public
  service.
\newblock \emph{Public Integrity}, 24\penalty0 (1):\penalty0 66--81, 2022.

\bibitem[Abdelmaboud et~al.(2022)Abdelmaboud, Ahmed, Abaker, Eisa, Albasheer,
  Ghorashi, and Karim]{abdelmaboud2022blockchain}
Abdelzahir Abdelmaboud, Abdelmuttlib Ibrahim~Abdalla Ahmed, Mohammed Abaker,
  Taiseer Abdalla~Elfadil Eisa, Hashim Albasheer, Sara~Abdelwahab Ghorashi, and
  Faten~Khalid Karim.
\newblock Blockchain for iot applications: taxonomy, platforms, recent
  advances, challenges and future research directions.
\newblock \emph{Electronics}, 11\penalty0 (4):\penalty0 630, 2022.

\bibitem[Leng et~al.(2022)Leng, Chen, Huang, Zhu, Su, Lin, and
  Zhang]{leng2022secure}
Jiewu Leng, Ziying Chen, Zhiqiang Huang, Xiaofeng Zhu, Hongye Su, Zisheng Lin,
  and Ding Zhang.
\newblock Secure blockchain middleware for decentralized iiot towards industry
  5.0: A review of architecture, enablers, challenges, and directions.
\newblock \emph{Machines}, 10\penalty0 (10):\penalty0 858, 2022.

\bibitem[Yaqoob et~al.()Yaqoob, Ahmed, Hashem, Ahmed, Gani, Imran, and
  Guizani]{yaqoob_internet_2017}
Ibrar Yaqoob, Ejaz Ahmed, Ibrahim Abaker~Targio Hashem, Abdelmuttlib
  Ibrahim~Abdalla Ahmed, Abdullah Gani, Muhammad Imran, and Mohsen Guizani.
\newblock Internet of things architecture: Recent advances, taxonomy,
  requirements, and open challenges.
\newblock 24\penalty0 (3):\penalty0 10--16.
\newblock ISSN 1536-1284.
\newblock \doi{10.1109/MWC.2017.1600421}.
\newblock URL \url{http://ieeexplore.ieee.org/document/7955906/}.

\bibitem[G{\"u}rpinar et~al.(2022)G{\"u}rpinar, Austerjost, Kamphues,
  Maa{\ss}en, Yildirim, and Henke]{gurpinar2022blockchain}
Tan G{\"u}rpinar, Maximilian Austerjost, Josef Kamphues, Jonas Maa{\ss}en,
  Furkan Yildirim, and Michael Henke.
\newblock Blockchain technology as the backbone of the internet of things--an
  introduction to blockchain devices.
\newblock In \emph{Proceedings of the Conference on Production Systems and
  Logistics: CPSL 2022}, pages 733--743. Hannover: publish-Ing., 2022.

\bibitem[Siegfried et~al.()Siegfried, Rosenthal, and
  Benlian]{siegfried_blockchain_2022}
Nils Siegfried, Tobias Rosenthal, and Alexander Benlian.
\newblock Blockchain and the industrial internet of things: A requirement
  taxonomy and systematic fit analysis.
\newblock 35\penalty0 (6):\penalty0 1454--1476.
\newblock ISSN 1741-0398.
\newblock \doi{10.1108/JEIM-06-2018-0140}.
\newblock URL
  \url{https://www.emerald.com/insight/content/doi/10.1108/JEIM-06-2018-0140/full/html}.

\bibitem[Nickerson et~al.(2013)Nickerson, Varshney, and
  Muntermann]{nickerson2013method}
Robert~C Nickerson, Upkar Varshney, and Jan Muntermann.
\newblock A method for taxonomy development and its application in information
  systems.
\newblock \emph{European Journal of Information Systems}, 22\penalty0
  (3):\penalty0 336--359, 2013.

\bibitem[Zwitter and Hazenberg(2020)]{zwitter2020decentralized}
Andrej Zwitter and Jilles Hazenberg.
\newblock Decentralized network governance: blockchain technology and the
  future of regulation.
\newblock \emph{Frontiers in Blockchain}, 3:\penalty0 12, 2020.

\bibitem[Psaras and Dias(2020)]{psaras2020interplanetary}
Yiannis Psaras and David Dias.
\newblock The interplanetary file system and the filecoin network.
\newblock In \emph{2020 50th Annual IEEE-IFIP International Conference on
  Dependable Systems and Networks-Supplemental Volume (DSN-S)}, pages 80--80.
  IEEE, 2020.

\bibitem[Fan et~al.(2023)Fan, Zhong, Guo, Chai, and Romano]{fan2023connecting}
Xinxin Fan, Zhi Zhong, Dong Guo, Qi~Chai, and Simone Romano.
\newblock Connecting smart devices to smart contracts with w3bstream.
\newblock In \emph{2023 IEEE International Conference on Blockchain and
  Cryptocurrency (ICBC)}, pages 1--2. IEEE, 2023.

\bibitem[Savolainen et~al.()Savolainen, Juslenius, Andrews, Pokrovskii,
  Tarkoma, and Pihkala]{savolainen2020streamr}
Petri Savolainen, Santeri Juslenius, Eric Andrews, Miroslav Pokrovskii, Sasu
  Tarkoma, and Henri Pihkala.
\newblock The streamr network: Performance and scalability.
\newblock \emph{url: https://streamrpublic. s3. amazonaws.
  com/streamr-network-scalability-whitepaper-2020-08-20. pdf}.

\bibitem[Lalley and Weyl(2018)]{lalley2018quadratic}
Steven~P Lalley and E~Glen Weyl.
\newblock Quadratic voting: How mechanism design can radicalize democracy.
\newblock In \emph{AEA Papers and Proceedings}, volume 108, pages 33--37.
  American Economic Association 2014 Broadway, Suite 305, Nashville, TN 37203,
  2018.

\bibitem[Ballandies et~al.()Ballandies, Dapp, and
  Pournaras]{ballandies_decrypting_2022}
Mark~C. Ballandies, Marcus~M. Dapp, and Evangelos Pournaras.
\newblock Decrypting distributed ledger design—taxonomy, classification and
  blockchain community evaluation.
\newblock 25\penalty0 (3):\penalty0 1817--1838.
\newblock ISSN 1386-7857, 1573-7543.
\newblock \doi{10.1007/s10586-021-03256-w}.
\newblock URL \url{https://link.springer.com/10.1007/s10586-021-03256-w}.

\bibitem[Hunhevicz et~al.(2021)Hunhevicz, Wang, Hess, and
  Hall]{hunhevicz2021no1s1}
Jens~Juri Hunhevicz, Hongyang Wang, Lukas Hess, and Daniel Hall.
\newblock no1s1--a blockchain-based dao prototype for autonomous space.
\newblock In \emph{Proceedings of the 2021 European Conference on Computing in
  Construction}, volume~2, pages 27--33. University College Dublin, 2021.

\end{thebibliography}

\end{document}